\documentclass{iopart}
\usepackage{iopams}
\usepackage{graphicx}
\usepackage{subfigure}
\usepackage{dcolumn}
\usepackage{bm}
\usepackage{cite}
\usepackage{color}

\begin{document}
\bibliographystyle{myiop}

\newcommand {\bra}[1]{\langle \: #1 \: |}
\newcommand {\ket}[1]{| \: #1 \: \rangle}
\newcommand {\unity}{{\textrm{1}\hspace*{-0.55ex}\textrm{l}}}
\newcommand {\dotp}[3]{\langle \: #1 | \: #2 \: | \: #3 \: \rangle}
\newcommand {\expect}[1]{\langle \: #1 \: \rangle}
\newcommand {\ybco}[1]{YBa$_2$Cu$_3$O$_{#1}$}
\newcommand {\ybcoE}{YBa$_2$Cu$_4$O$_{8}$ }
\newcommand {\ybcoEnd}{YBa$_2$Cu$_4$O$_{8}$}
\newcommand {\lacuo}{La$_2$CuO$_4$ }
\newcommand {\lasco}{La$_{2-x}$Sr$_{x}$CuO$_{4}$ }
\newcommand {\cuo}{CuO$_2$}
\newcommand {\Kbar}[3]{{^{#1}\overline{K}_{#2}^{#3}}}
\newcommand {\Ks}[2]{{K_{#1}^{#2}}}
\newcommand {\Ksn}[1]{{K_{01}^{#1}}}
\newcommand {\tone}{{^{k}\!T_{1 \alpha}^{-1}}}
\newcommand {\us}[2]{{^{#1}U_{#2}}}
\newcommand {\Vs}[2]{{^{#1}V_{#2}}}
\newcommand {\Vsn}[2]{{^{#1}V_{#2}^{0}}}
\newcommand {\Vfc}[2]{{^{#1}V_{#2}^{\mathit{fc}}}}
\newcommand {\Tm}[2]{{^{#1}T_{1#2}^{-1}}}
\newcommand {\Tone}[2]{ {^{#1}T}_{1#2}}
\newcommand {\usbar}[2]{{^{#1}\overline{U}_{#2}}}
\newcommand {\ustil}[2]{{^{#1}\tilde{U}_{#2}}}
\newcommand {\TTm}[2]{\left({^{#1}T}_{1#2}T\right)^{-1}}
\newcommand {\gam}[1]{{^{#1}\gamma}}
\newcommand {\Q}[3]{{^{#1}V_{#2}^{#3}}}
\newcommand {\taus}[2]{{^{#1}\tau}_{#2}}
\newcommand {\taueff}{$\tau_{\mathit{eff}}$}
\newcommand {\taueffm}{\tau_{\mathit{eff}}}
\newcommand {\Tg}[2]{{^{#1}T_{2G#2}^{-1}}}
\newcommand {\chir}[1]{\chi_{#1}'}
\newcommand {\Tgind}{T_{2G,\mathrm{ind}}^{-1}}

\title[Anisotropy of the AFM spin correlations in the superconducting state of YBCO]{Anisotropy of the antiferromagnetic spin correlations in the superconducting state of \ybco{7} and \ybcoE}

\author{A Uldry, M Mali, J Roos and P F Meier}

\address{Physics Institute, University of Zurich, CH-8057 Zurich, 
         Switzerland}

\eads{\mailto{uldryac@pauli.physik.unizh.ch,\mailto{pfmeier@physik.unizh.ch}}}

\begin{abstract} 
We present evidence that the antiferromagnetic spin correlations in optimally doped \ybco{7} and underdoped \ybcoE develop a surprisingly strong anisotropy in the superconducting state. Comparing the ratio of the nuclear spin-lattice relaxation rates of the planar copper and oxygen, measured at the lowest and highest temperatures as well as at $T_c$, we conclude that the antiferromagnetic in-plane correlations vanish as the temperature goes to zero. This observation is corroborated by the measurement of the copper linewidth in \ybcoEnd. In contrast, the out-of-plane correlations do not change appreciably between $T=T_c$ and $T=0$. Within a model of fluctuating fields this extreme anisotropy of the antiferromagnetic correlations also explains the observed temperature dependence of the anisotropy of the copper relaxation measured in a low external magnetic field.
\end{abstract}




Nuclear magnetic/quadrupole resonance (NMR/NQR) experiments have shed light on 
the antiferromagnetic (AFM) spin correlations in the normal state of high-temperature superconductors. It was found that the correlations become progressively stronger as the temperature is reduced towards the superconducting transition temperature $T_c$. The question is to what extent these correlations persist below $T_c$. As the temperature is lowered in the superconducting state, the spin-lattice relaxation rate $\Tm{k}{\alpha}$ of the planar copper ($k=63$) and planar oxygen ($k=17$) nuclei decrease rapidly, with the applied field either parallel to the CuO$_2$ plane ($\alpha=ab$) or perpendicular to it ($\alpha=c$). This decrease goes approximately like $T^3$, which is the temperature dependence expected for d-wave orbital pairing~\cite{mp90}. Great emphasis has been put on the different temperature behaviour of the copper and oxygen relaxation rates in the normal state. However, less attention has been payed on the ratios of these rates~\cite{hammel:89,taki:91,yosh92,borsa:92,bankay:92,martind:93,tomeno:94,bankay:94} in the superconducting state. A difficulty arises since in the normal state, the spin-lattice relaxation is largely insensitive to the strength of the applied magnetic field~\cite{mitro:02,zheng99}, while the relaxation rate becomes field dependent in the superconducting state at low temperature, with $\Tm{63}{c}$ showing a stronger dependence than $\Tm{63}{ab}$ and $\Tm{17}{c}$~\cite{borsa:92,martind:92,bankay:92,martind:93}. In order to draw any conclusions about magnetism in the superconducting state it is therefore very important to look for intrinsic effects which can only be obtained from experiments done in weak magnetic fields so as to minimise the flux line influence. We will consider four NMR/NQR experimental results: the ratio $\Tm{63}{c}/\Tm{17}{c}$, the NQR copper linewidth in \ybcoEnd, the nuclear spin-spin relaxation rate $\Tg{}{}$ and the ratio $\Tm{63}{ab}/\Tm{63}{c}$. From these experiments we conclude that in the superconducting state the AFM in-plane correlations vanish as the temperature goes to zero, whereas the out-of-plane correlations do not change much between $T=T_c$ and $T=0$. 
In the case of the first three experiments the discussion is model-independent. For the fourth the argumentation is based on a model of fluctuating fields developed recently~\cite{uldryU}. For convenience and clarity, we have however adopted throughout the paper the formalism and notation of this model which we outline below.\\

Some of us have recently proposed a new phenomenological model for the analysis of nuclear spin-lattice relaxation rate experiments in the normal state of the cuprates~\cite{uldryU}. Special attention was payed to the question of to what degree the hyperfine fields originating from magnetic moments on copper ions should be added coherently. That the question is of importance is demonstrated by the anisotropy ratio~\footnote{The experimental value for the anisotropy $\Tm{63}{ab}/\Tm{63}{c}$ in the normal state of \ybco{7} is significantly higher than what is calculated assuming either a fully coherent or a fully incoherent addition of the hyperfine fields. Therefore the experimental result can only be explained if the model provides an interpolation scheme which exhibits a maximum between these two extremes.} $\Tm{63}{ab}/\Tm{63}{c}$. The in-plane and out-of-plane AFM correlations 
could be determined in~\cite{uldryU} for the optimally doped and underdoped compound of the 
\ybco{y} family. The temperature dependences of the in-plane and out-of-plane correlations were found to be similar in the normal state. In the present report we extend the application of the model to the analysis of data obtained in the superconducting 
state, in particular to the question of the various measured ratios of the relaxation 
rates. In the model of fluctuating fields~\cite{book:slichter} the NMR spin-lattice relaxation rate of a nuclear species $k$ is determined by fluctuating fields in the direction $\beta$ and $\gamma$ perpendicular to the applied static field direction $\alpha$ and is thus expressed as 
\begin{equation}\label{defTm}
  \Tm{k}{\alpha}(T)=\lbrack \Vs{k}{\beta}(T)+\Vs{k}{\gamma}(T)\rbrack 
\taueffm(T).
\end{equation}
The term $\taueffm (T)$ is an effective electronic spin-spin correlation time, and $\Vs{k}{\beta}(T)$ and $\Vs{k}{\gamma}(T)$ correspond to the square of the components $\beta$ and $\gamma$ of the effective hyperfine fields at the nucleus. In the cuprates the hyperfine fields are produced by more or less localised electronic moments on the copper ions. In particular, for a planar oxygen one gets~\cite{uldryU} 
\begin{equation}\label{VsdefO}
  \Vs{17}{\beta}(T) =\;\frac{1}{4 \hbar^2}\; 2 \; C^2_{\beta}
\lbrack 1 + \Ksn{\beta}(T)  \rbrack,  
\end{equation}
where $C_\beta$ is the hyperfine field which is transferred from the two Cu moments (at site $0$ and $1$) adjacent to the O. $\Ksn{\beta}$ is the $\beta$-component of the normalised nearest-neighbour electron spin-spin correlation defined as $\Ksn{\beta}(T)=4\langle S_0^\beta S_1^\beta \rangle$ and can take values between $-1$ (fully antiferromagnetically correlated and yielding $\Vs{17}{\beta}=0$) and $0$ (no correlation). The antiferromagnetism observed in the parent compounds is well described by a nearly-isotropic two-dimensional Heisenberg model ($J^c\approx J^{ab}$). Upon doping, the long range order is destroyed but a short range order persists, characterised by spin-spin correlation lengths $\lambda^c$ and $\lambda^{ab}$ of the order of a lattice constant. In Ref. \cite{uldryU} the spin-spin correlations were parameterised according to $\Ksn{\beta}(T)=-\mathrm{exp}[-1/\lambda^{\beta}(T)]$. Typical values for \ybco{7} at $T_c$ (as determined in \cite{uldryU}) are $\Ksn{ab}(T=T_c)=-0.4$ and $\Ksn{c}(T=T_c)=-0.5$. It should be emphasised that they are static correlations with respect to typical NMR times.\\
In contrast to oxygen, a copper nucleus is affected by an on-site anisotropic field ($A_\beta$) and by transfered isotropic hyperfine fields ($B$) originating from its 4 nearest neighbour copper ions. This leads to an expression for $\Vs{63}{\beta}(T)$ that contains further distant spin correlations as well. For simplicity however, these further distant correlations have been assumed to depend on $\Ksn{\beta}$ and to decrease exponentially with 
the distance between spins. As a result $\Vs{63}{\beta}(T)$ can be 
expressed as 
\begin{equation}\label{VsdefCu}
 \fl  \Vs{63}{\beta}(T) = \;\frac{1}{4 \hbar^2}\; \lbrack A^2_{\beta} + 
4 B^2 + 8 A_{\beta} B \Ksn{\beta}(T) \nonumber 
  + 8  B^2 |\Ksn{\beta}(T)|^{\sqrt{2}} + 4 B^2 
|\Ksn{\beta}(T)|^{2} \rbrack.
\end{equation}

\begin{figure}
\begin{center}
\includegraphics*[width=8cm]{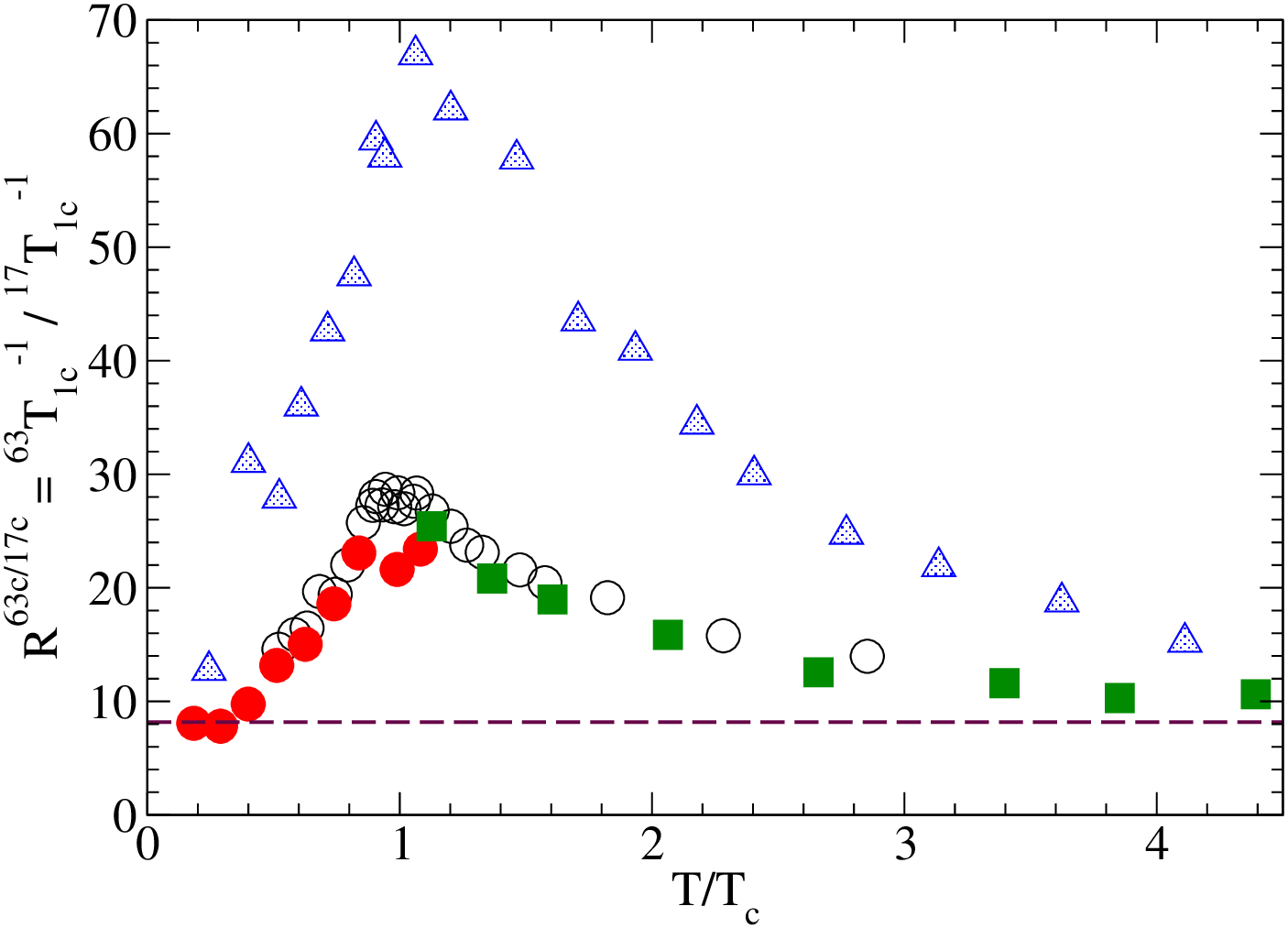}
\caption{$R^{63c/17c}(T)=\Tm{63}{c}/\Tm{17}{c}$ versus $T/T_c$ for \ybco{6.96} in high field (empty circles, data from Yoshinari \etal~\cite{yosh92}) and for \ybco{7} in low field (filled circles, data from Martindale \etal~\cite{martind:93}). The squares denote values calculated from combining $\Tm{63}{c}$ data from Barrett \etal~\cite{barret:91} and $\Tm{16}{c}$ data from Nandor \etal~\cite{nandor:99}. The ratio for \ybcoE (half-filled triangles) was combined in high field for the oxygen measurement only, from the data from 
Bankay \etal~\cite{bankay:94}). The dashed line is the model prediction for vanishing AFM correlations ($\Ksn{ab}=0$).\label{WCuO}}
\end{center}
\end{figure}

Valuable information on AFM spin correlations is gained from ratios of relaxation rates, since then $\taueffm$ [Eq. (\ref{defTm})] cancels out. We gather in Fig. \ref{WCuO} some experimental data for $R^{63c/17c}:=\Tm{63}{c}/\Tm{17}{c}$ in \ybco{7} (circles \cite{yosh92,martind:93}, squares \cite{barret:91,nandor:99}) and \ybcoE (triangles \cite{bankay:94}). 
Note that the ratio for \ybco{7} at elevated temperature has been obtained by combining copper data from Barrett \etal~\cite{barret:91} and oxygen measurements from Nandor \etal~\cite{nandor:99}.
Both optimally and underdoped compounds see an increase of $R^{63c/17c}$ from high temperature down to $T_c$, followed by a sharp decrease as the temperature is further reduced down to the very same level that is observed at the highest temperature. It is astonishing that this striking similarity of $R^{63c/17c}$ at very low and very high temperatures has not been much interpreted so far. One reason for this may be that in the standard analysis of $\Tm{k}{\alpha}$ data (the so-called MMP theory~\cite{mmp}), the copper and oxygen are treated very differently. In the model outlined in Eqs. (\ref{defTm}-\ref{VsdefCu}) however, the relaxation rate of both nuclei is determined by the same $\taueffm$. It is only $\Vs{63}{\beta}$ and $\Vs{17}{\beta}$ that differ due to the different contribution of hyperfine field values and spin correlations. \\
The normal state temperature dependence of $\Tm{63}{c}$ and $\Tm{17}{c}$ could be fitted very well with the model (\ref{defTm}-\ref{VsdefCu}), whereby their ratio is given by
\begin{equation}\label{R6317}
R^{63c/17c}(T)=\frac{2 \Vs{63}{ab}(T)}{\Vs{17}{a}(T)+\Vs{17}{b}(T)}.
\end{equation}
We note that the temperature dependence of $R^{63c/17c}$ comes solely from $\Ksn{ab}(T)$, the in-plane AFM correlations. At very high temperature we expect all AFM correlations to go to zero. In such a case (\ref{R6317}) reduces to
\begin{equation}\label{R6317:0}
R_0^{63c/17c}=\frac{A_{ab}^2+4B^2}{C_a^2+C_b^2}.
\end{equation}
Using the hyperfine constants calculated in \cite{uldryU} (in units of $10^{-6}$ eV: $A_{ab}=0.168$, $B=0.438$, $C_a=0.259$, $C_b=0.173$), we find that $R_0^{63c/17c}=8.2$, a value marked by the dashed line in Fig. \ref{WCuO}. It is obvious from Fig. \ref{WCuO} that this is also, to a very good agreement, the experimental $T\rightarrow 0$ limit of this ratio. Therefore we conclude that the in-plane correlations reduce to zero in the superconducting state. It has of course been recognised in earlier works that the significant decrease in $\Tm{63}{c}/\Tm{17}{c}$ in the superconducting state suggests a loss of AFM fluctuations~\cite{tomeno:94}. However, the connection with the measurements 
at elevated temperature had not been made. We would like to emphasise that Eq. (\ref{R6317:0}) is a widely accepted result in case of no correlations and is quite independent of our phenomenological model. In particular, Eq. (\ref{R6317:0}) will result from any model that explains spin-lattice relaxation rates in terms of fluctuating hyperfine fields added 
incoherently. \\

Further indication that the in-plane AFM correlations decrease below $T_c$ is given by the temperature dependence of the copper NQR linewidth of YBa$_2$Cu$_4$O$_{8}$. 
\begin{figure}
\begin{center}
\includegraphics*[width=8cm]{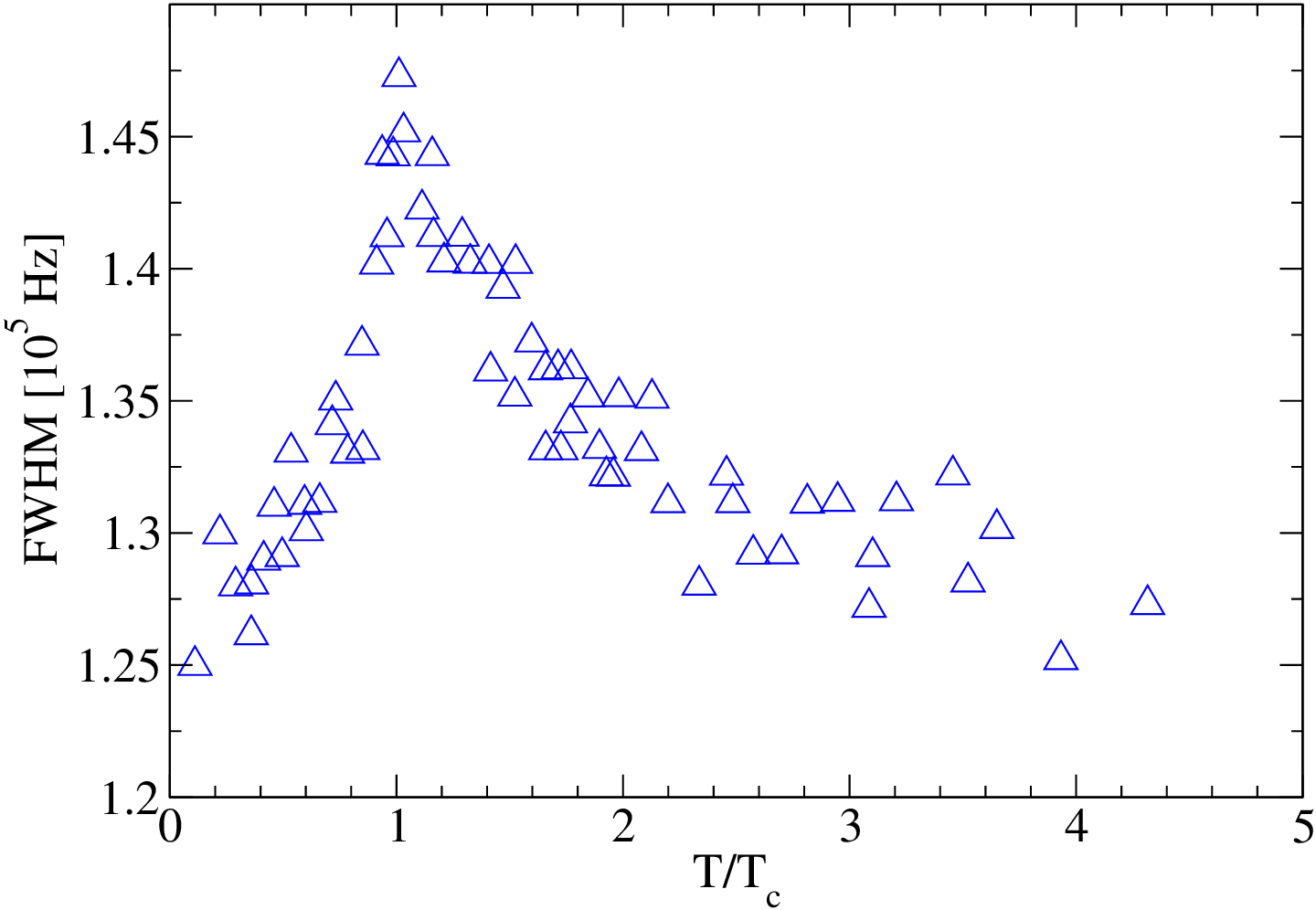}
\caption{Linewidth versus $T/T_c$ for \ybcoEnd. Data from Mali \etal~\cite{mali:02}. \label{LW}}
\end{center}
\end{figure}
This underdoped compound has the advantage of being stoichiometric with a well ordered and stable structure, without the otherwise inevitable disorder effects created by extrinsic dopands. The temperature dependence of the planar copper linewidth from~\cite{mali:02} is reproduced in  Fig. \ref{LW}. The measurements were made on a loose polycrystalline powder sample. The observed linewidth results from two contributions: a large temperature independent quadrupolar contribution and a smaller temperature dependent magnetic 
contribution. As seen in Fig. \ref{LW}, the linewidth increases with decreasing temperature down to $T_c$, very much like $R^{63c/17c}$, and upon entering the superconducting state also decreases sharply. The exact origin of the temperature dependent magnetic contribution is at the moment not known precisely. However, the two main sources of the magnetic line broadening are the static local magnetic fields stemming from the material-imperfection induced staggered magnetisation, and the indirect nuclear spin-spin interactions mediated by the electron spin system in the plane. In contrast to NMR where the static magnetic line broadening is predominantly caused by the local magnetic field components {\it parallel} to the large applied magnetic field, the NQR lines are broadened mainly by local magnetic fields that are {\it perpendicular} to the NQR quantisation 
axis~\cite{abragam}. This axis is material specific and is fixed by the largest principal axis of the electric field gradient (EFG) tensor. At the plane-copper site in \ybcoE the largest principal axis of EFG tensor coincides with the crystallographic c-axis. 
Therefore any decrease of the in-plane magnetic fields due to the loss of in-plane AFM correlations will result, as observed in the experiment, in a decrease of the magnetic contribution to the linewidth of the plane-copper NQR line.\\

\begin{figure}
\begin{center}
\includegraphics*[width=8cm]{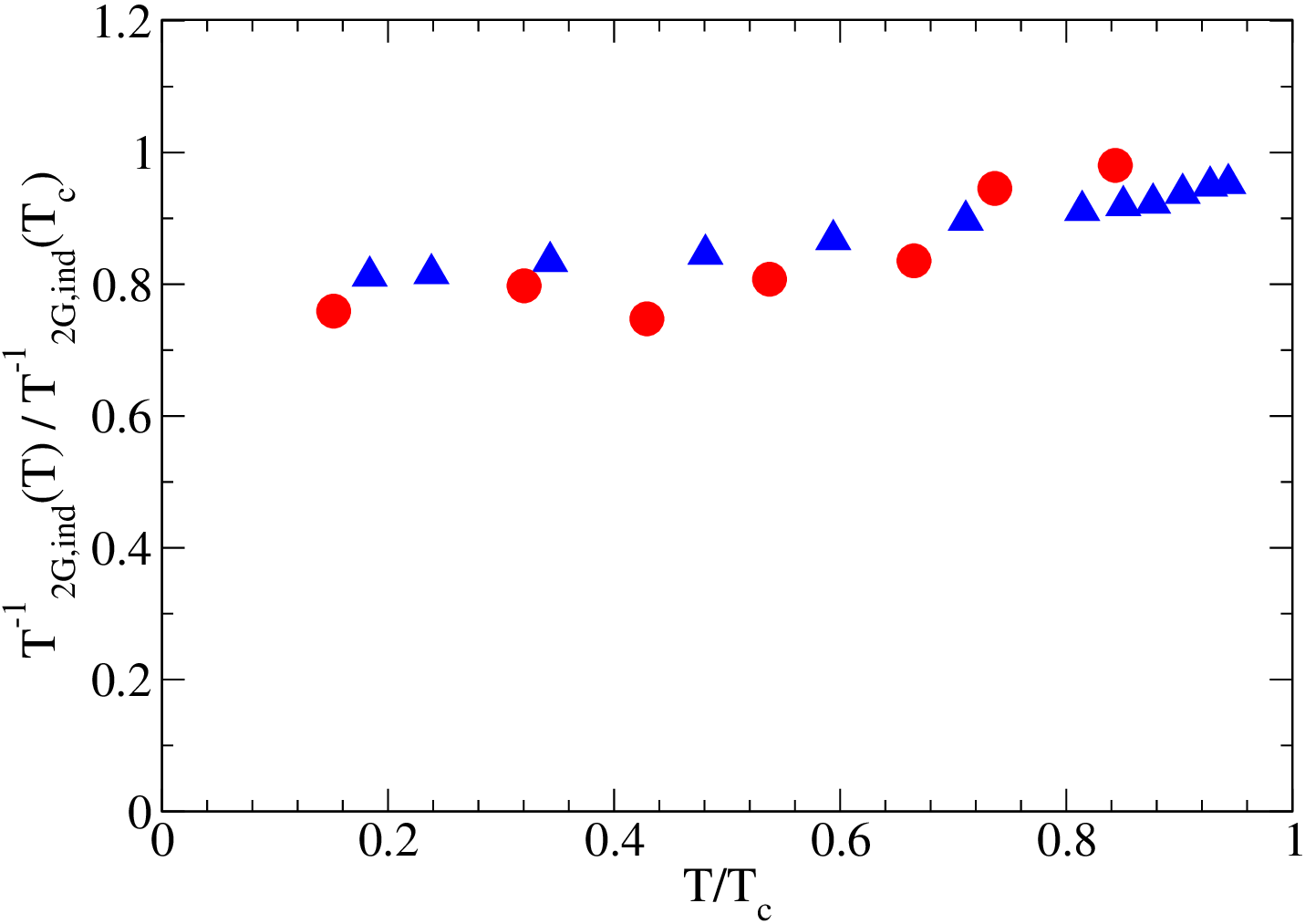}
\caption{$\Tgind(T)/\Tgind(T_c)$ versus $T/T_c$ for \ybco{7} (circles) and \ybcoE (triangles). Data from Stern \etal~\cite{stern:95}. \label{T2G}}
\end{center}
\end{figure}

We turn now to an experiment that provides information on the component of the correlations along the c-axis. It has been known from $\Tg{}{}$ measurements that AFM correlations do subsist in the superconducting state almost as strong as in the normal state. Fig. \ref{T2G} reproduces a plot from Stern \etal~\cite{stern:95} showing normalised NQR measurements of $T^{-1}_{2G,\mathrm{ind}}$, the Gaussian contribution to the nuclear spin-spin relaxation $T^{-1}_{2G}$ caused by the indirect nuclear spin-spin coupling mediated by the non-local static spin susceptibility.
However, in the \ybco{y} compounds, $\Tgind(T)$ depends only on the component of real part of the electronic spin susceptibility along the c-axis~\cite{haase:99}, and hence it depends only on the correlations along this axis. In our notation therefore, $\Tgind(T)$ depends on $\Ks{01}{c}$ and not on $\Ks{01}{ab}$. The measurements on \ybco{7} (circles \cite{tsh:91}) and \ybcoE (triangles \cite{bankay:92}) reported in Fig. \ref{T2G} indicate that the out-of-plane AFM correlations vary little from their value at $T_c$ when the temperature is lowered. This suggests that whereas the in-plane AFM correlations vanish in the superconducting state, the out-of-plane correlations remain more or less 
frozen in.\\

\begin{figure}
\begin{center}
\includegraphics*[width=8cm]{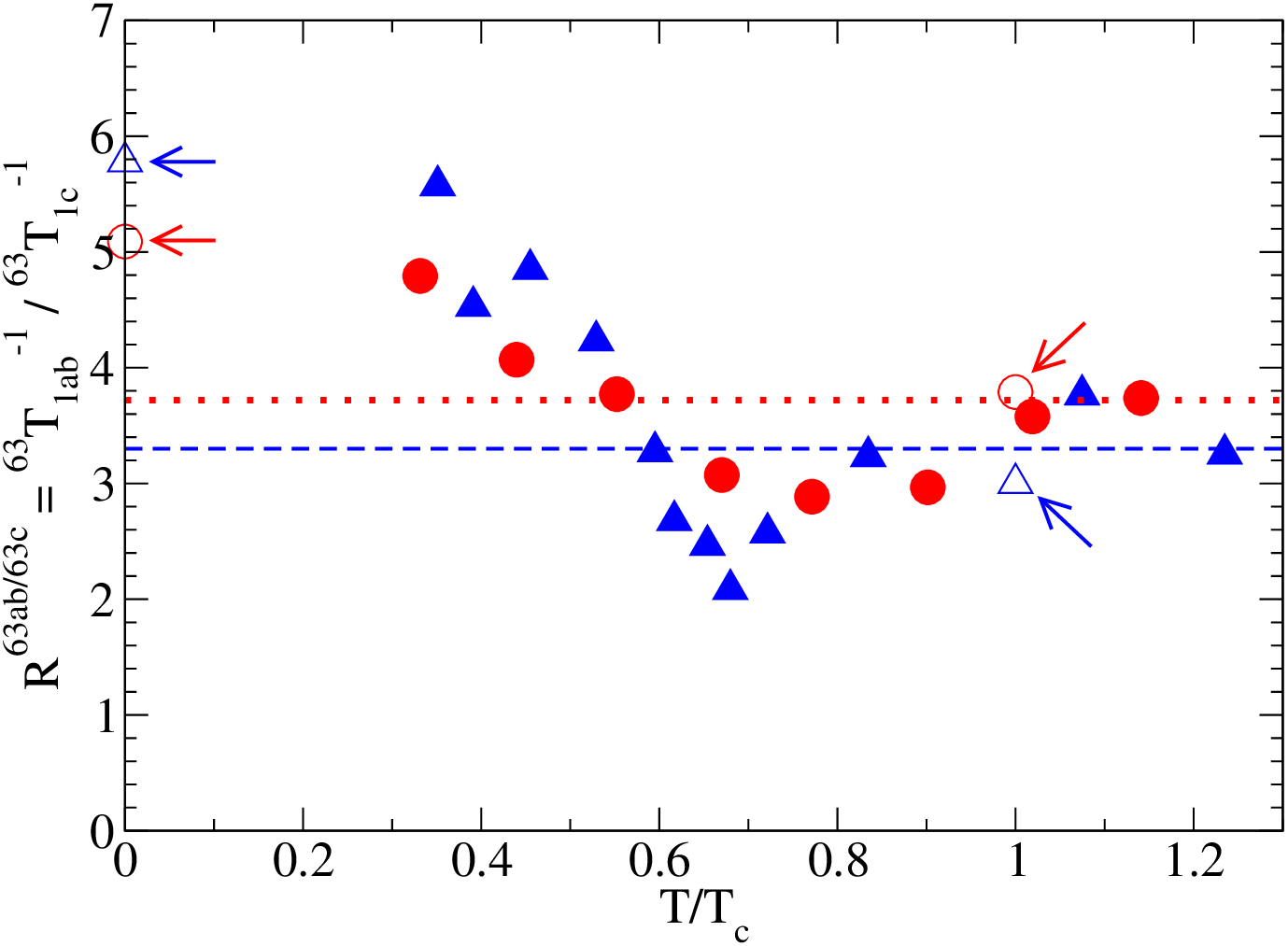}
\caption{$\Tm{63}{ab}/\Tm{63}{c}$ versus $T/T_c$ for \ybco{7} (filled circles, data from Takigawa \etal~\cite{tsh:91}) and \ybcoE (filled triangles, data from Bankay \etal~\cite{bankay:92}) in low field. The dotted (\ybco{7}) and dashed lines (\ybcoE) are the normal state measurements. The empty symbols are model predictions. \label{WCuCu}}
\end{center}
\end{figure}

Finally, we check these drastically different temperature dependencies for the in-and out-of-plane components of the correlation on experiments where both components are involved. The development of such an extreme anisotropy of the AFM correlations in the superconducting state has visible consequences for the planar copper relaxation rate anisotropy $R^{63ab/63c}:=\Tm{63}{ab}/\Tm{63}{c}$. Fig. \ref{WCuCu}, reproduced from Bankay \etal~\cite{bankay:92}, shows the temperature dependence of $R^{63ab/63c}$ in \ybco{7} (filled circles~\cite{tsh:91}) and \ybcoE (filled triangles~\cite{bankay:92}). 
Both ratios were measured in a weak magnetic field. In the normal state, the anisotropy is temperature-independent and larger for \ybco{7} (dotted line) than for \ybcoE (dashed line) \cite{zimm}. After entering the superconducting state an upturn occurs, so that at low temperature the anisotropy of \ybcoE exceeds that of \ybco{7}.\\
Within the framework of the phenomenological model Eqs. (\ref{defTm}-\ref{VsdefCu}), the anisotropy is expressed as follows
\begin{equation}\label{R6363}
R^{63ab/63c}(T)=\frac{1}{2}\left(1
+\frac{\Vs{63}{c}(T)}{\Vs{63}{ab}(T)} \right).
\end{equation}
According to the suggestion that the out-of-plane correlations are frozen at their $T_c$ value in the superconducting state and the in-plane ones have dropped to zero at $T=0$, $R^{63ab/63c}$ at $T=0$ becomes
\begin{equation}\label{R6363:0}
R_0^{63ab/63c}=\frac{1}{2}\left(1+\frac{\Vs{63}{c}[\Ks{01}{c}(T=T_c)] }{\Vs{63}{ab}[\Ks{01}{ab}=0]} \right).
\end{equation}
In order to compute $R^{63ab/63c}(T_c)$ and $R_0^{63ab/63c}$ we need to know, besides the hyperfine field constants, the values at $T_c$ of the in-plane correlation $\Ks{01}{ab}(T=T_c)$ and of the out-of-plane correlations $\Ks{01}{c}(T=T_c)$. For \ybco{7} we take the values determined in \cite{uldryU}, $\Ks{01}{c}(T=T_c)=-0.50$ and $\Ks{01}{ab}(T=T_c)=-0.40$, and get $R^{63ab/63c}(T_c)=3.79$ and $R_0^{63ab/63c}=5.09$. These results are marked by empty circles at $T=T_c$ and $T=0$ in the Fig. \ref{WCuCu}. In the case of \ybcoE we do not have values for $\Ks{01}{\alpha}(T=T_c)$, but we can use the results found in \cite{uldryU} for the underdoped compound \ybco{6.63}, whose planar charge carrier concentration comes close to that of YBa$_2$Cu$_4$O$_{8}$. Therefore, taking $[\Ks{01}{c}(T=T_c)]=-0.61$ and $[\Ks{01}{ab}(T=T_c)]=-0.53$, we get for \ybcoE $R^{63ab/63c}(T_c)=3.00$ and $R_0^{63ab/63c}=5.78$. These results are marked by empty triangles at $T=T_c$ and $T=0$ in the Fig. \ref{WCuCu}~\footnote{In Eq. \ref{R6363:0}, the value of $\Vs{63}{ab}[\Ks{01}{ab}=0]$ is the same for any compound with the same hyperfine field constants. However, $\Vs{63}{c}[\Ks{01}{c}(T=T_c)]$ is higher for \ybcoE than for \ybco{7}, since the AFM correlations 
at $T_c$ 
are higher in the underdoped than in the optimally doped compounds. Hence $R_0^{63ab/63c}$ is higher in the former than in the latter.}. For both compounds the model predictions reproduce well the limits of the temperature behaviour of this ratio. \\

In conclusion, we have shown that a range of NMR/NQR experiments indicates that below $T_c$, the AFM spin correlations develop a different behaviour in the in-plane direction and in the out-of-plane direction. From the analysis of $R^{63c/17c}$ and the planar copper NQR linewidth in \ybcoE we deduced that the AFM in-plane correlations disappear gradually in the superconducting state, in contrast to the out-of-plane correlations which remain almost unchanged, as demonstrated by the measurements of the planar copper spin-spin relaxation $\Tg{}{}$. An additional evidence that an extreme anisotropy of the AFM correlation develops in the superconducting state is provided by the analysis of $R^{63ab/63c}$ within the model of fluctuating fields. The predicted values at $T=0$ come astonishingly close to the experiment. Expressed in terms of the AFM correlation lengths $\lambda^c$ and $\lambda^{ab}$, we find that $\lambda^{ab}$ vanishes as $T\rightarrow 0$. This means that the short range correlations of the spin system, which above $T_c$ retained the nearly-isotropic Heisenberg-like character of the parent antiferromagnet, acquire below $T_c$ with decreasing temperature a more and more Ising-like character. We finally point out that these conclusions are drawn from NMR data, which are sensitive to the quasiparticles spectrum at very low energies. It is possible that neutron scattering data taken at higher energy might give a different picture. \\

We thank E. Stoll, S. Renold, T. Mayer and C. Bersier for interesting discussions, and C. P. Slichter for his interest and encouragements. This work was carried under the auspices of the Swiss National Science Foundation.\\


\bibliography{mybib}

\begin{thebibliography}{10}

\bibitem{mp90}
Monien H and Pines D 1990 {\em Phys. Rev. B} {\bf 41} 6297.

\bibitem{hammel:89}
Hammel P~C, Takigawa M, Heffner R~H, Fisk Z and Ott K~C 1989 {\em Phys. Rev.
  Lett.} {\bf 63} 1992.

\bibitem{taki:91}
Takigawa M, Reyes A~P, Hammel P~C, Thompson J~D, Heffner R~H, Fisk Z and Ott
  K~C 1991 {\em Phys. Rev. B} {\bf 43} 247.

\bibitem{yosh92}
Yoshinari Y, Yasuoka H and Ueda Y 1992 {\em J. Phys. Soc. Japan} {\bf 61} 770.

\bibitem{borsa:92}
Borsa F, Rigamonti A, Corti M, Ziolo J, Hyun Oi-Bae and Torgeson D~R 1992 {\em
  Phys. Rev. Lett.} {\bf 68} 698.

\bibitem{bankay:92}
Bankay M, Mali M, Roos J, Mangelschots I and Brinkmann D 1992 {\em Phys. Rev.
  B} {\bf 46} R11228.

\bibitem{martind:93}
Martindale J~A, Barrett S~E, O'Hara K~E, Slichter C~P, Lee W~C and Ginsberg D~M
  1993 {\em Phys. Rev. B} {\bf 47} R9155.

\bibitem{tomeno:94}
Tomeno I, Machi T, Tai K, Koshizuka N, Kambe S, Hayashi A, Ueda Y and Yasuoka H
  1994 {\em Phys. Rev. B} {\bf 49} 15327.

\bibitem{bankay:94}
Bankay M, Mali M, Roos J and Brinkmann D 1994 {\em Phys. Rev. B} {\bf 50} 6416.

\bibitem{mitro:02}
Mitrovi\'c V~F, Bachman H~N, Halperin W~P, Reyes A~P, Kuhns P and Moulton W~G
  2002 {\em Phys. Rev. B} {\bf 66} 014511.

\bibitem{zheng99}
Zheng G-Q, Clark W~G, Kitaoka Y, Asayama K, Kodama Y, Kuhns P and Moulton W~G
  1999 {\em Phys. Rev. B} {\bf 60} R9947.

\bibitem{martind:92}
Martindale J~A, Barrett S~E, Klug C~A, O'Hara K~E, DeSoto S~M, Slichter C~P,
  Friedmann T~A and Ginsberg D~M 1992 {\em Phys. Rev. Lett.} {\bf 68} 702.

\bibitem{uldryU}
Uldry A and Meier P~F 2005 Analysis of {NMR} {S}pin-{L}attice {R}elaxation
  {R}ates in {C}uprates Accepted for publication in {\it Phys. Rev. B}, {\it
  Preprint} cond-mat/0502075.

\bibitem{book:slichter}
Slichter C~P 1996 {\em Principles of Magnetic Resonance} (Berlin: Sprin\-ger).

\bibitem{barret:91}
Barrett S~E, Martindale J~A, Durand D~J, Pennington C~H, Slichter C~P,
  Friedmann T~A, Rice J~P and Ginsberg D~M 1991 {\em Phys. Rev. Lett.} {\bf 66}
  108.

\bibitem{nandor:99}
Nandor V~A, Martindale J~A, Groves R~W, Vyaselev O~M, Pennington C~H, Hults L
  and Smith J~L 1999 {\em Phys. Rev. B} {\bf 60} 6907.

\bibitem{mmp}
Millis A~J, Monien H and Pines D 1990 {\em Phys. Rev. B} {\bf 42} 167.

\bibitem{mali:02}
Mali M, Roos J, Keller H, Karpinski J and Conder K 2002 {\em Phys. Rev. B} {\bf
  65} 184518.

\bibitem{abragam}
Abragam A 1986 {\em The Principles of Nuclear Magnetism} (Oxford: Clarendon
  Press) chapter VII, page 254.

\bibitem{stern:95}
Stern R, Mali M, Roos J and Brinkmann D 1995 {\em Phys. Rev. B} {\bf 51} 15478.

\bibitem{haase:99}
Haase J, Morr D~K and Slichter C~P 1999 {\em Phys. Rev. B} {\bf 59} 7191.

\bibitem{tsh:91}
Takigawa M, Smith J~L and Hults W~L 1991 {\em Phys. Rev. B} {\bf 44} R7764.

\bibitem{zimm}
Zimmermann H, Mali M, Bankay M and Brinkmann D 1991 {\em Physica C} {\bf
  185-189} 1145.

\end{thebibliography}

\end{document}